\documentclass[conference]{IEEEtran}
\IEEEoverridecommandlockouts
\usepackage{cite}
\usepackage{amsmath,amssymb,amsfonts}
\usepackage{algorithmic}
\usepackage{graphicx}
\usepackage{textcomp}
\usepackage{xcolor}
\usepackage{multirow}
\def\BibTeX{{\rm B\kern-.05em{\sc i\kern-.025em b}\kern-.08em
    T\kern-.1667em\lower.7ex\hbox{E}\kern-.125emX}}
\begin{document}

\title{SEGSRNet for Stereo-Endoscopic Image Super-Resolution and Surgical Instrument Segmentation\\

}

\author{\IEEEauthorblockN{Mansoor Hayat}
\IEEEauthorblockA{\textit{Dept. of Electrical Engineering} \\
\textit{Chulalongkorn University}\\
Bangkok, Thailand \\
6471015721@student.chula.ac.th}
\and
\IEEEauthorblockN{Supavadee Aramvith}
\IEEEauthorblockA{\textit{Dept. of Electrical Engineering} \\
\textit{Chulalongkorn University}\\
Bangkok, Thailand \\
supavadee.a@chula.ac.th}
\and
\IEEEauthorblockN{Titipat Achakulvisut}
\IEEEauthorblockA{\textit{Dept. of Biomedical Engineering} \\
\textit{Mahidol University}\\
Bangkok, Thailand \\
titipat.ach@mahidol.edu}
}
\maketitle

\begin{abstract}
SEGSRNet addresses the challenge of precisely identifying surgical instruments in low-resolution stereo endoscopic images, a common issue in medical imaging and robotic surgery. Our innovative framework enhances image clarity and segmentation accuracy by applying state-of-the-art super-resolution techniques before segmentation. This ensures higher-quality inputs for more precise segmentation. SEGSRNet combines advanced feature extraction and attention mechanisms with spatial processing to sharpen image details, which is significant for accurate tool identification in medical images. Our proposed model, SEGSRNet, surpasses existing models in evaluation metrics including PSNR and SSIM for super-resolution tasks, as well as IoU and Dice Score for segmentation. SEGSRNet can provide image resolution and precise segmentation which can significantly enhance surgical accuracy and patient care outcomes.

\end{abstract}


\section{Introduction}

The advancement of digital imaging technology, from early monochromatic photography to modern 8k resolution, plays a pivotal role in various fields, including medical diagnostics, where image clarity is essential \cite{b1}. In medical imaging, high-resolution techniques are crucial, particularly in diagnostics and surgical procedures, underscoring the importance of super-resolution (SR) techniques to overcome issues like lens limitations \cite{b3} \cite{b33}. 

In stereo image SR, maintaining view consistency is vital, with recent developments like the Parallax Attention Module in Disparity Constraint Stereo SR (DCSSR) \cite{b3} and bi Directional Parallax Attention Map (biPAM) in iPASSR \cite{b16} enhancing perceptual quality. Accurate identification and segmentation of surgical instruments in images are important, for which advanced semantic segmentation techniques are employed, leveraging CNNs and architectures like U-Net \cite{b10} for improved accuracy.

Our research integrates SR and segmentation technologies for robotic-assisted surgeries. We introduce a hybrid model that applies SR before segmentation, enhancing the accuracy with high-quality inputs. This model, benchmarked against established methods like UNet \cite{b10} and TernausNet \cite{b15}, shows superior performance in both SR and segmentation domains, demonstrating its efficacy in complex medical imaging tasks.

\begin{figure*}[t]
\centering
\includegraphics[width=0.8\textwidth]{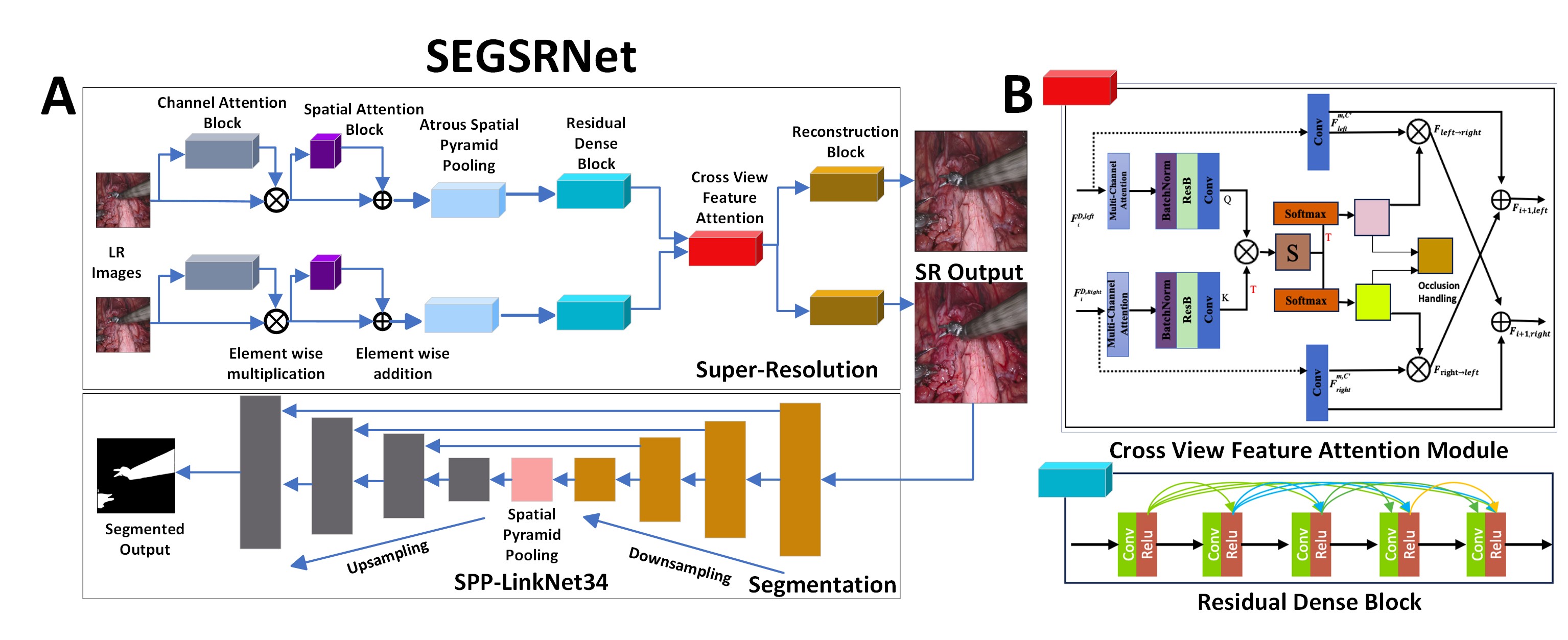}
\caption{Proposed SEGSRNet Architecture. A. SEGSRNet architecture consists of super-resolution and segmentation modules. B. Proposed cross-view attention module and residual dense block in super-resolution framework. }
\label{fig1}
\end{figure*}

\begin{figure*}[t]
\centering
\includegraphics[width=0.8\textwidth]{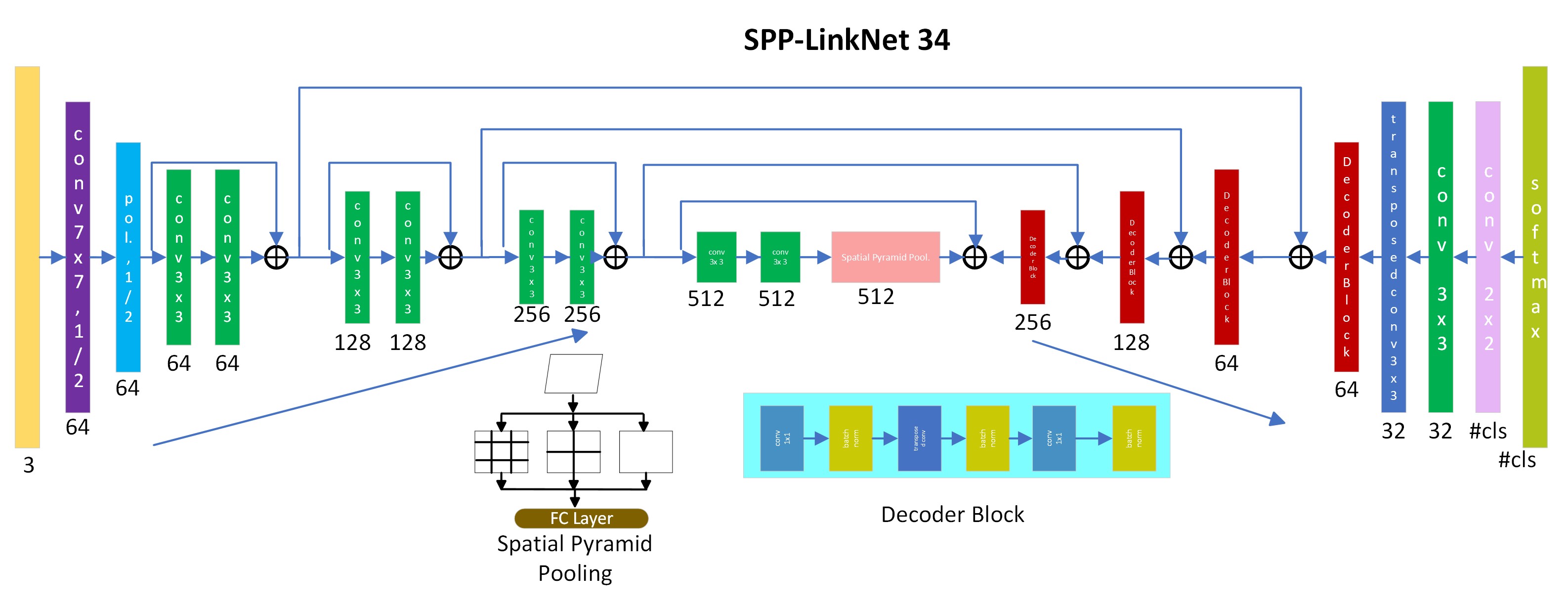}
\caption{SPP-LinkNet34 structure highlighting the encoder-decoder network with spatial pyramid pooling for enhanced multi-scale feature extraction in image segmentation tasks.}
\label{fig}
\end{figure*}
\section{Research Methodology}
Advancements in super-resolution (SR) techniques, especially the biPAM network, have significantly improved surgeons' sensory capabilities in medical settings. Demonstrating its effectiveness in the NTIRE 2022 Challenge \cite{b17}, biPAM excels in learning cross-view information, which is pivotal for high-quality SR stereo images. This process involves downscaling high-resolution (HR) images to create low-resolution (LR) counterparts, which are then enhanced through a feature extraction module comprising a combined channel and spatial attention (CCSA) and an Atrous Spatial Pyramid Pooling (ASPP) block, followed by Residual Dense Blocks (RDB). The network culminates in SR image reconstruction, leveraging multi-attention biPAM.
\begin{figure*}[t]
\centering
\includegraphics[width=0.8\textwidth]{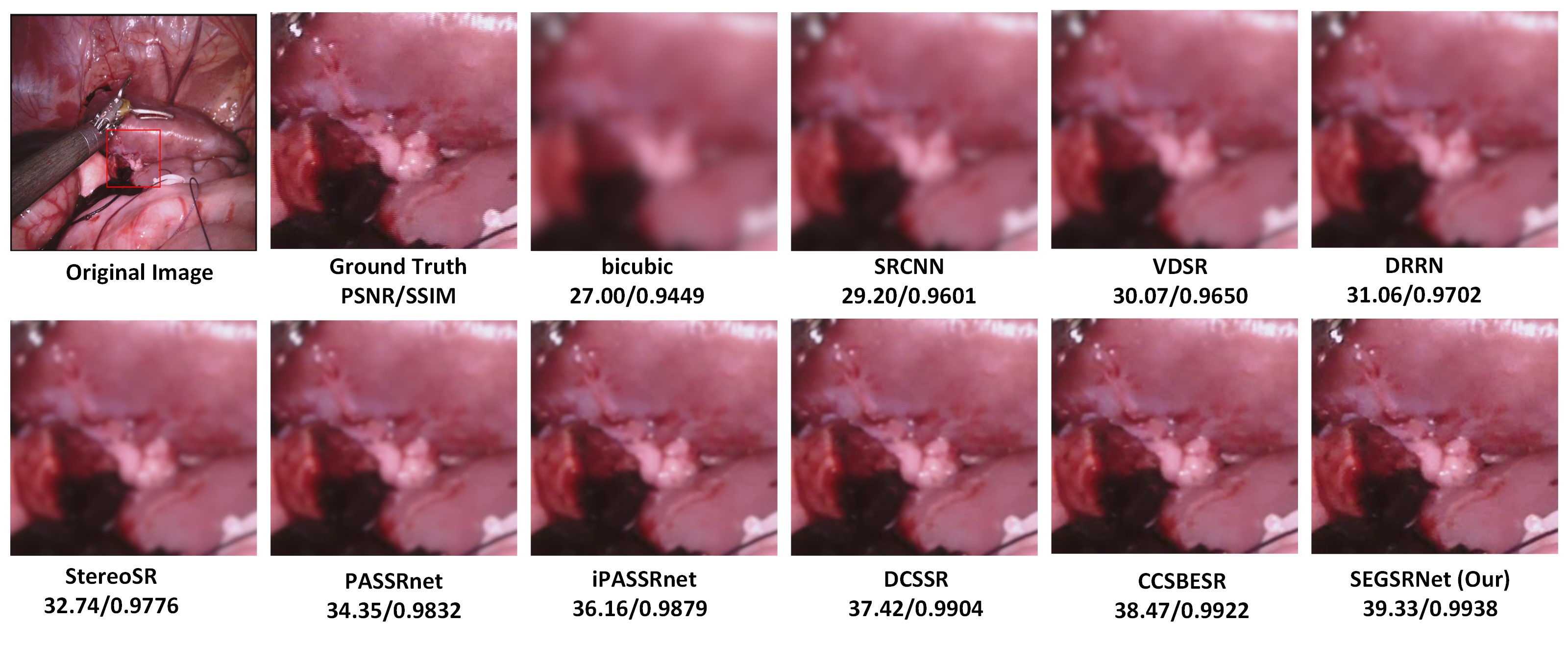}
\caption{Assessment of the Visual Quality of High-Resolution Images Created Through Image Super-Resolution Techniques at a $\times4$ Scale Factor.}
\label{fig3}
\end{figure*}
Our proposed network addresses the challenge of semantic segmentation in surgery by distinguishing surgical instruments from the background in super-resolution images, enhancing both medical imaging and robotic surgery.

\subsection{Super-Resolution Part}
\subsubsection{Feature Extraction and Refinement Blocks}
Our model features a Combined Channel and Spatial Attention Block (CCSB)\cite{b18}, which includes a Channel Attention Block (CAB) for enhancing feature maps and a Spatial Channel Attention Block (SAB) for focusing on key regions. The features processed through CCSB are further refined using an Atrous Spatial Pyramid Pooling (ASPP) block and Residual Dense Blocks (RDBs). These components deepen feature extraction and create a comprehensive feature hierarchy, significantly improving the model's performance in super-resolution.

\subsubsection{Cross-View Feature Interaction Module}
Integrating multi-scale attention into biPAM enhances the interaction and integration of cross-view information in stereo features, which is key for precise stereo correspondence. This improvement is achieved through hierarchical feature representation by combining output features from each Residual Dense Block (RDB) in the feature extraction module \cite{b32}.

Inputs to biPAM undergo processing through batch normalization and a transition residual block (ResB), followed by 1×1 convolutions, producing feature tensors \( FU \) and \( FV \):

\begin{equation}
F_X(h,w,c) = F_X(h,w,c) - \frac{1}{W} \sum_{i=1}^{W} F_X(h,i,c) \quad \text{for } X \in {U, V}
\end{equation}

The multi-scale attention mechanism enhances stereo-image processing by adaptively focusing on and integrating details from different resolution levels, cardinal for accurately reconstructing depth information. Attention maps \( M_{R \rightarrow L} \) and \( M_{L \rightarrow R} \) facilitate cross-view interaction:

\begin{equation}
F_{X \rightarrow Y} = M_{X \rightarrow Y} \otimes F_X \quad \text{for } (X,Y) \in \{(R,L)\}.
\end{equation}

The occlusion handling scheme computes valid masks \( V_L \) and \( V_R \), ensuring continuous spatial distributions by filling occluded regions with features from the target view:

\begin{equation}
F_{X \rightarrow Y} = V_Y \cdot F_{Y \rightarrow X} + (1 - V_Y) \cdot F_Y \quad \text{for } (X,Y) \in \{(R,L)\}.
\end{equation}

This module significantly enhances stereo image processing by effectively improving feature interaction and managing occlusions.

\subsubsection{Reconstruction Block}
In our model's reconstruction block, a refinement block combines \( F_{R \rightarrow L} \) with \( F_L \), followed by processing through a Residual Dense Block (RDB) and a channel attention layer (CALayer). This sequence, including additional RDBs, convolution layers, and a sub-pixel layer, significantly enhances feature fusion and image quality, leading to a high-precision, super-resolved image.

\subsection{Segmentation Part}
Binary segmentation differentiates between foreground and background, parts segmentation identifies individual components of objects, and type segmentation classifies each pixel based on object categories, enhancing scene comprehension and object interaction analysis. SPP-LinkNet-34, depicted in Fig. 2, features an architecture optimized for effective segmentation with an encoder-decoder structure. It employs convolution techniques, batch normalization, and ReLU non-linearity \cite{b20}, \cite{b21}. The encoder utilizes a 7×7 kernel and spatial max-pooling, followed by residual blocks \cite{b22}, while the decoder is designed for efficient feature mapping.

A notable aspect of SPP-LinkNet-34 is its use of the lighter ResNet18 as its encoder \cite{b19}, and the inclusion of a Spatial Pyramid Pooling (SPP) block that enhances multi-scale input handling. This design allows SPP-LinkNet-34 to recover spatial information lost during downsampling efficiently, resulting in improved segmentation accuracy and efficiency, suitable for real-time applications.
\subsection{Datasets}
We use The MICCAI 2018 Robotic Scene Segmentation Sub-Challenge \cite{b23} ("MICCAI 2018") and the MICCAI 2017 Robotic Instrument Segmentation Challenge \cite{b24} ("EndoVis 2017") are  the datasets for the robotic scene segmentation in endoscopic procedures. Both challenges provide high-resolution stereo image pairs \(1280 \times 1024\) pixels) captured during porcine training sessions, along with  camera calibrations. For the SR task, we trained our models using the MICCAI 2018 dataset and then evaluated their performance using both the MICCAI 2018 and EndoVis 2017 datasets. Conversely, for segmentation, we conducted both training and testing exclusively on the EndoVis 2017 dataset. We calculate Peak signal-to-noise ratio (PSNR) and Structural Similarity (SSIM) as evaluation metrics. For segmentation performace, we use 10-folds cross-validation to measure segmentation  Intersection over Union (IoU) on EndoVis 2017 and calculate the mean and standard deviation (STD) of the validation set.
\subsection{Training Settings}
Our proposed model was implemented using the Pytorch 2.0 framework and trained on a  Nvidia 3090Ti GPU. We employed Xavier initialization for the model's parameters and used Adam optimizer with an initial learning rate of \(3 \times 10^{-4}\). All models are trained for 100 epochs. We used a batch size of 6 for images at scale two ($\times2$) and a batch size of 5 for images at scale four ($\times4$).

\section{Experimental Results}

\begin{figure*}[t]
\centering
\includegraphics[width=0.8\textwidth]{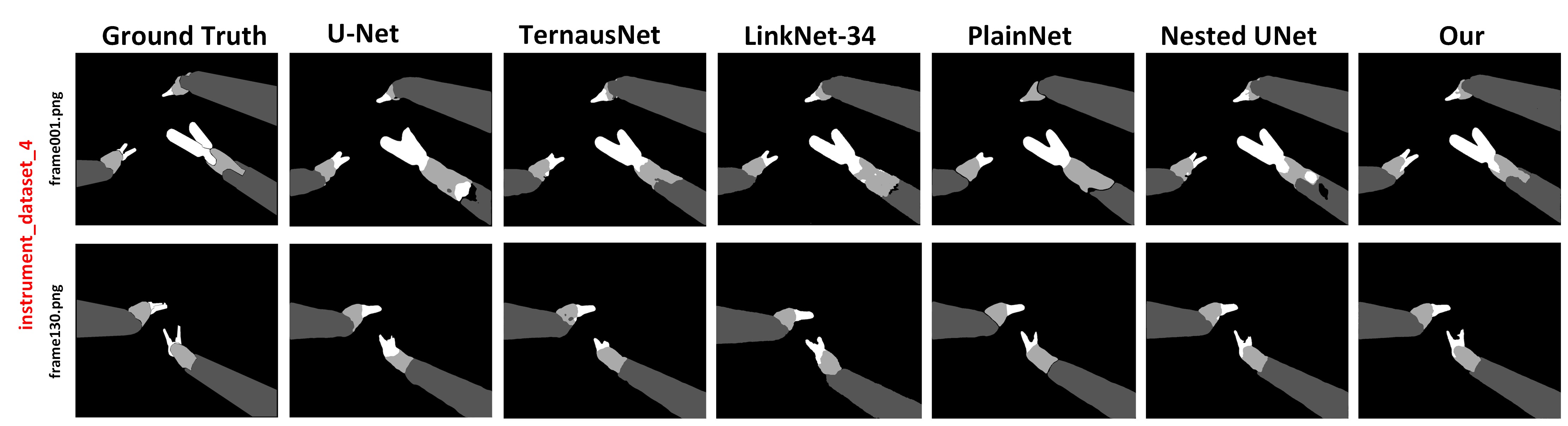}
\caption{Comparative Evaluation of Segmentation Performance: Our Model versus Current State-of-the-Art Models.}
\label{fig:seg_com}
\end{figure*}

\begin{table}[h]
\centering
\caption{\textbf{Performance Evaluation of Enlargement Factors $\times2$ And $\times4$ On MICCAI 2018 And EndoVis 2017: A Quantitative Analysis Using PSNR/SSIM.}}
\label{table}
\setlength{\tabcolsep}{3pt}
\begin{tabular}{|p{60pt}|p{20pt}|p{70pt}|p{70pt}|}
\hline
\textbf{Method} & \textbf{Scale} & \textbf{MICCAI 2018 \newline(PSNR/SSIM) $\uparrow$} & \textbf{EndoVis 2017 \newline(PSNR/SSIM) $\uparrow$} \\ 
\hline
bicubic & $\times2$ & 38.60/0.9792 & 27.07/0.9594 \\ \hline
SRCNN & $\times2$ & 38.99/0.9811 & 28.89/0.9646 \\ \hline
VDSR & $\times2$ & 39.57/0.9824 & 29.23/0.9654 \\ \hline
DRRN & $\times2$ & 40.18/0.9858 & 32.19/0.9666 \\ \hline
StereoSR & $\times2$ & 40.25/0.9859 & 36.18/0.9912 \\ \hline
PASSRNet & $\times2$ & 40.36/0.9860 & 40.36/0.9921 \\ \hline
iPASSRNet & $\times2$ & 41.01/0.9866 & 40.57/0.9941 \\ \hline
DCSSRNet  & $\times2$ & 41.09/0.9866 & 40.03/0.9917 \\ \hline
CCSBESR  & $\times2$ & 41.99/0.9871 & 40.38/0.9920 \\ \hline
SEGSRNet (Our) & $\times2$ & \textbf{42.41/0.9879} & \textbf{41.87/0.9965}  \\ \hline
\multicolumn{3}{}{} \\ \hline
\textbf{Method} & \textbf{Scale} & \textbf{MICCAI 2018 \newline(PSNR/SSIM) $\uparrow$} & \textbf{EndoVis 2017 \newline(PSNR/SSIM) $\uparrow$} \\ 
\hline
bicubic  & $\times4$ & 32.85/0.9480 & 25.72/0.9436 \\ \hline
SRCNN  & $\times4$ & 33.11/0.9510 & 26.71/0.9513 \\ \hline
VDSR  & $\times4$ & 33.35/0.9516 & 27.06/0.9518 \\ \hline
DRRN & $\times4$ & 34.01/0.9558 & 28.79/0.9624 \\ \hline
StereoSR  & $\times4$ & 34.08/0.9545 & 34.04/0.9669 \\ \hline
PASSRNet & $\times4$ & 34.12/0.9547 & 36.83/0.9699 \\ \hline
iPASSRNet  & $\times4$ & 34.52/0.9549 & 37.76/0.9710 \\ \hline
DCSSRNet  & $\times4$ & 34.76/0.9553 & 33.52/0.9719 \\ \hline
CCSBESR  & $\times4$ & 34.99/0.9558 & 37.91/0.9725 \\ \hline
SEGSRNet (Our) & $\times4$ & \textbf{36.01/0.9768}  & \textbf{38.33/0.9924}  \\ \hline
\end{tabular}
\end{table}

\begin{table*}[!htbp]
\centering
\caption{Analysis of Segmentation Performance for Instruments Across Three Different Tasks (Mean ± Standard Deviation)}
\begin{tabular}{|l|c|c|c|c|c|c|}
\hline
\multirow{2}{*}{\textbf{Methods}} & \multicolumn{2}{c|}{\textbf{Binary segmentation}} & \multicolumn{2}{c|}{\textbf{Parts segmentation}} & \multicolumn{2}{c|}{\textbf{Type segmentation}} \\ \cline{2-7}
 & \textbf{IoU(\%) $\uparrow$} & \textbf{Dice(\%) $\uparrow$} & \textbf{IoU(\%) $\uparrow$} & \textbf{Dice(\%) $\uparrow$} & \textbf{IoU(\%) $\uparrow$} & \textbf{Dice(\%) $\uparrow$} \\ \hline
U-Net  & 75.44 ± 18.18 & 84.37 ± 14.58 & 48.41 ± 17.59 & 60.75 ± 18.21 & 15.80 ± 15.06 & 23.59 ± 19.87 \\ \hline
TernausNet & 81.14 ± 19.11 & 88.07 ± 14.63 & 62.23 ± 16.48 & 74.25 ± 15.55 & 34.61 ± 20.53 & 45.86 ± 23.20 \\ \hline
LinkNet-34 & 82.36 ± 18.77 & 88.87 ± 14.35 & 34.55 ± 20.96 & 41.26 ± 23.44 & 22.47 ± 35.73 & 24.71 ± 37.54 \\ \hline
PlainNet & 81.86 ± 15.85 & 88.96 ± 12.98 & 64.73 ± 17.39 & 73.53 ± 16.98 & 34.57 ± 21.93 & 44.64 ± 25.16 \\ \hline
Nested UNet & 82.94 ± 16.82 & 89.42 ± 14.01 & 58.38 ± 19.06 & 69.59 ± 18.66 & \textbf{41.72 ± 33.44} & \textbf{48.22 ± 34.46} \\ \hline
\textbf{SPP-LinkNet34 (Our)}  & \textbf{83.65 ± 16.47}  & \textbf{89.80 ± 13.99}  & \textbf{66.87 ± 17.10}  & \textbf{76.93 ± 16.08} & 15.96 ± 13.78 & 23.79 ± 18.88 \\ \hline
\end{tabular}
\end{table*}

\subsection{Quantitative and Qualitative Results}
Our model outperforms traditional U-Net by 9.81\% and 27.60\% in binary and parts segmentation in terms of IoU. However, our model had limitations in type segmentation due to its emphasis on global contextual information which is less suited for the fine-grained, pixel-level distinctions required among multiple complex classes, compared to the more generalized tasks in binary and parts segmentation.

The super-resolution results show that SEGSRNet corrects inaccuracies and computes disparities more effectively than traditional methods such as Bicubic interpolation and DRRN (Fig 3). Overall, SEGSRNet outperforms traditional model in both $\times2$ and $\times4$. After applying SR, we illustrates the model’s proficiency in various segmentation tasks, including binary, Parts, and Type segmentation, highlighting its superior performance in accurately segmenting different image components (Fig 4).

\section{Conclusion}
SRSEGNet introduces a breakthrough in deep learning for super-resolution and segmentation in endoscopic vision, leveraging convolutional neural networks and SPP-LinkNet-34. Achieving high performance on the EndoVis 2017 dataset, SRSEGNet excels in binary segmentation with an IoU of 83.65\% and a Dice score of 89.80\%, effectively handling complex multi-class segmentation tasks.

\vspace{12pt}

\end{document}